\documentclass[prl,twocolumn,superscriptaddress,floats]{revtex4}

\usepackage[dvips]{graphicx}
\usepackage{amsmath}
\usepackage{booktabs}
\usepackage{dcolumn}

\usepackage{rotating}

\newlength{\La} 

\newlength{\Lb} 

\newlength{\Lc} 

\newcolumntype{d}{D{.}{.}{-1}}

\newcommand{\grad}{\ensuremath{^\circ}}

\newcommand{\tn}{T$_N$}

\newcommand{\tzg}{t$_{2g}$}

\newcommand{\muB}{$\mu_B$}

\newcommand{\latio}{LaTiO$_3$}

\begin{document}

\advance\vsize by 2 cm

\title{Crystal and magnetic structure of \latio :
evidence for non-degenerate $t_{2g}$-orbitals}

\author{M.~ Cwik }
\affiliation{II. Physikalisches Institut, Universit\"at zu K\"oln,
Z\"ulpicher Str. 77, D-50937 K\"oln, Germany}

\author{T. Lorenz}
\affiliation{II. Physikalisches Institut, Universit\"at zu K\"oln,
Z\"ulpicher Str. 77, D-50937 K\"oln, Germany}

\author{J. Baier}
\affiliation{II. Physikalisches Institut, Universit\"at zu K\"oln,
Z\"ulpicher Str. 77, D-50937 K\"oln, Germany}

\author{R. M\"uller}
\affiliation{II. Physikalisches Institut, Universit\"at zu K\"oln,
Z\"ulpicher Str. 77, D-50937 K\"oln, Germany}

\author{G. Andr\'e}
\affiliation{Laboratoire L\'eon Brillouin,
C.E.A./C.N.R.S., F-91191 Gif-sur-Yvette CEDEX,
France}

\author{F. Bour\'ee}
\affiliation{Laboratoire L\'eon Brillouin,
C.E.A./C.N.R.S., F-91191 Gif-sur-Yvette CEDEX,
France}

\author{F. Lichtenberg}
\affiliation{Experimentalphysik VI, EKM, Institut f\"ur Physik,
Universit\"at Augsburg, 86135 Augsburg, Germany}

\author{A. Freimuth}
\affiliation{II. Physikalisches Institut, Universit\"at zu K\"oln,
Z\"ulpicher Str. 77, D-50937 K\"oln, Germany}

\author{R. Schmitz}
\affiliation{Institut f\"ur Theoretische Physik, 
Universit\"at zu K\"oln,
Z\"ulpicher Str. 77, D-50937 K\"oln, Germany}

\author{E. M\"uller-Hartmann}
\affiliation{Institut f\"ur Theoretische Physik, 
Universit\"at zu K\"oln,
Z\"ulpicher Str. 77, D-50937 K\"oln, Germany}

\author{M.~ Braden}
\affiliation{II. Physikalisches Institut, Universit\"at zu K\"oln,
Z\"ulpicher Str. 77, D-50937 K\"oln, Germany}

\date{\today, \textbf{DRAFT}}

\pacs{PACS numbers:}

\begin{abstract}

The crystal and magnetic structure of \latio ~ has been
studied by x-ray and neutron diffraction techniques
using nearly stoichiometric samples.
We find a strong structural anomaly near the
antiferromagnetic ordering, T$_N$=146\ K.
In addition, the octahedra in \latio ~ exhibit
an intrinsic distortion which implies a splitting
of the \tzg -levels. 
Our results indicate that \latio ~ should be considered 
as a Jahn-Teller system where the structural distortion and
the resulting level splitting are enhanced by the magnetic 
ordering.

\end{abstract}

\maketitle

\latio ~ has been studied already in the seventies and was thought
to be a text book example of a Mott-insulator with antiferromagnetic
order \cite{imada}.
Ti is in its three-valent state with a single electron
in the \tzg -orbitals of the 3d-shell. The titanate is hence an
electron analog to
the cuprates with a single hole in the 3d-shell. However, the
\tzg -orbitals in the \latio ~ are less Jahn-Teller active and,
therefore, the orbital moment may not be  fully quenched in the titanate.
The physics of the orbital degree of freedom has recently reattracted
attention to this material \cite{keimer,khaliullin}.

The ordered moment in \latio ~ amounts to 0.46 \muB ~ which is much
smaller than the value of  1\muB ~ expected for a single electron with
quenched orbital moment \cite{meijer}.
Quantum fluctuations can explain 
only about 15\% reduction in the 3D-case.
A straight-forward explanation could be given in terms of
spin-orbit coupling, as an unquenched orbital moment would 
align antiparallel to the spin-moment in the titanate.
However, in a recent neutron scattering
experiment the magnon spin gap was observed at 3.3\ meV, and
it was argued that
the strong interaction of an orbital moment with the crystal
lattice implies a much larger value for the spin gap \cite{keimer}.
An orbital contribution to the ordered moment in \latio ~
was hence excluded.
On the basis of standard theories, however, 
even the G-type antiferromagnetic ordering in \latio ~
may not be explained without a spin-orbit coupling.
Instead one expects ferromagnetism \cite{5,6} related with the 
orbital degeneracy.
Under the assumption of a specific
structural distortion, Moshizuki and Imada recently
presented a successful model for the antiferromagnetic order 
in \latio ~\cite{moshizuki}.
However, there is no experimental evidence for such a distortion.
The puzzling magnetic properties of \latio ~ led 
Khaliullin and Maekawa to suggest a novel theoretical description 
for RETiO$_3$ based on the idea of an orbital liquid. 
They were able to explain many of the
magnetic characteristics of LaTiO$_3$ \cite{khaliullin},
but the presumed orbital fluctuations have not been observed
\cite{fritsch}.
Therefore, magnetism in \latio ~still remains an open issue.

We have reanalyzed the crystal and magnetic structure of \latio ~
by x-ray and by neutron diffraction samples with almost perfect
stoichiometry.
First, we find a clear structural anomaly at the N\'eel-ordering
and, second, the shape of the octahedra in this compound is not ideal 
but distorted.
From these observations we conclude that \latio ~ has to be considered as 
a soft Jahn-Teller system thereby explaining many of its unusual 
magnetic properties.

Samples of \latio ~ were prepared following the standard solid state
reaction; however the last process was performed in a floating zone
image furnace. This procedure gave large single crystals with
small mosaic spread.
Since twinning would induce severe problems in any single
crystal diffraction experiment,
we crushed parts of these single crystals in order
to perform powder diffraction studies. Special care was attracted
to the stoichiometry of these samples. \latio ~has the tendency to
incorporate La and Ti vacancies which are frequently denoted by an excess
of oxygen in the formula LaTiO$_{3+\delta}$.
By adjusting the ratio of TiO and TiO$_2$ in the starting
materials \cite{lichtenberg} one may vary the final oxygen content of the samples.
Stoichiometry was verified by thermogravimetric analyzes and by determining
the N\'eel-temperature in a SQUID magnetometer.
For the stoichiometric sample studied
here we obtain \tn =146\ K, amongst the highest values reported so far
for \latio .
The temperature dependence of the lattice parameters was
determined on a Siemens D5000 x-ray diffractometer using 
Cu-K$_\alpha$ radiation.
Powder neutron diffraction experiments were performed at the
Orph\'ee-reactor in Saclay, using the high-resolution diffractometer
3T.2 ($\lambda$=1.2\AA) and the high-flux diffractometer G4.1
($\lambda$=2.4\AA).

\begin{table}

\begin{tabular}{|c|c|c|c|c|c|c|}

\hline
 \multicolumn{4}{|c|}{\rule[-3mm]{0mm}{8mm} \latio , $T_N =
150 K$}&\latio&YTiO$_3$\\\hline
&\multicolumn{5}{c|}{temperature (K)}\\\cline{2-6}
&  293 & 155 & 8 & 298 & 293\\\hline
 a &5.6336(1)&5.6353(1)&5.6435(1)&5.6247(1)&5.316(2)\\
 b &5.6156(1)&5.6021(1)&5.5885(1)&5.6071(1)&5.679(2)\\
 c &7.9145(2)&7.9057(2)&7.9006(2)&7.9175(1)&7.611(3)\\\hline
Ti-O2&2.057(1)&2.054(1)&2.053(1)&2.044(1)&2.077(4)\\
Ti-O2&2.031(1)&2.032(1)&2.032(1)&2.032(1)&2.016(4)\\
Ti-O1&2.0298(4)&2.0273(1)&2.0280(4)&2.028(1)&2.023(2)\\
O2-O2&2.935(2)&2.938(2)&2.944(2)&2.924&2.908(5)\\
O2-O2&2.845(2)&2.840(2)&2.832(2)&2.839&2.881(5)\\
\hline
x$_{RE}$&0.9916(3)&0.9930(3)&0.9930(3)&0.9929(3)&0.9793(1)\\
y$_{RE}$&0.0457(2)&0.0482(2)&0.0491(2)&0.0428(1)&0.0729(1)\\
z$_{RE}$&0.25       &0.25       &0.25&0.25&0.25\\

x$_{O1}$&0.0799(3)&0.0799(3)&0.0813(3)&0.0781(3)&0.121(1)\\
y$_{O1}$&0.4913(3)&0.4939(3)&0.4940(3)&0.4904(3)&0.458(1)\\
z$_{O1}$&0.25&0.25&0.25&0.25&0.25\\

x$_{O2}$&0.7096(3)&0.7087(3)&0.7092(3)&0.7104(2)&0.691(1)\\
y$_{O2}$&0.2941(3)&0.2946(3)&0.2943(3)&0.2914(2)&0.310(1)\\
z$_{O2}$&0.0417(1)&0.0421(1)&0.0428(1)&0.0412(1)&0.058(1)\\
\hline
U$_{La}$&0.0062(2)&0.0052(2)&0.0013(2)&0.0162(3)&0.0055(4)\\
U$_{Ti}$&0.0036(4)&0.0047(4)&0.0008(4)&0.0104(5)&0.0042(6)\\
U$_{O1}$&0.0075(3)&0.0066(3)&0.0040(3)&0.0174(4)&0.006(3)\\
U$_{O2}$&0.0069(2)&0.0064(2)&0.0034(2)&0.0183(3)&0.007(2)\\
\hline
$\Theta$ (\grad)&12.88(5)&12.86(5)&13.11(5)&11.26(4)&19.9(2)\\
$\Phi$ (\grad)&9.53(1)&9.69(1)&9.60(1)&8.08(1)&13.33(4)\\
\hline

\end{tabular}

\caption{Results of the neutron high resolution powder diffraction studies at
 different
temperatures, the two right columns present data from references
\cite{eitel} and \cite{maclean} respectively. Lattice constants are given in
\AA ~ and thermal parameters in \AA$^2$.}
\end{table}

At room temperature, \latio ~ exhibits the crystal structure of
GdFeO$_3$-type (space group $Pbnm$), see Fig. 1. This structure
arises from the ideal perovskite (space group $Pm\bar{3}m$) by 
tilting the TiO$_6$-octahedron around a [110]$_c$-axis (the subscript $c$
denotes the notation with respect to the
cubic perovskite lattice) followed by a rotation around the
c-axis (or [001]$_c$ axis). 
One may note that while
tilt distortions are opposite for any neighboring pair of
octahedra, octahedra neighboring along the c-axis
rotate in the same sense.

\begin{figure}[t]

\includegraphics*[width=0.73\columnwidth]{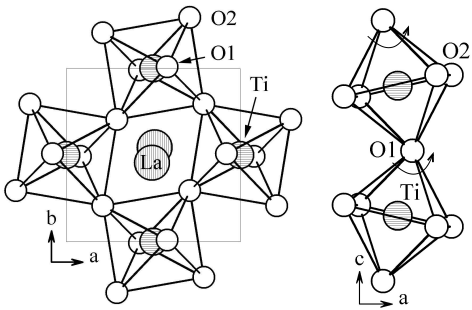}%

\caption{  Crystal structure of \latio ~ in space group $Pbnm$,
O2 denotes the oxygen in the $a,b$-planes O1 the apical-one.
The left part shows four
octahedra connected in the  $a,b$-plane, the right part
a pair of neighboring octahedra along $c$.
}
\label{fig:1}
\end{figure}

Fig. 2 shows the temperature dependence of the
orthorhombic lattice parameters and that of the crystal
volume. There are clear anisotropic anomalies
near the antiferromagnetic ordering.
Upon passing into the ordered state the lattice elongates
along $a$ and shrinks along $b$, whereas for the c-direction
and for the lattice volume no anomalous changes can be resolved.
Anomalies in the thermal expansion have also been observed
independently in a capacitance dilatometer experiment  \cite{bruns}.

\begin{figure}[t]
\includegraphics*[width=0.75\columnwidth]{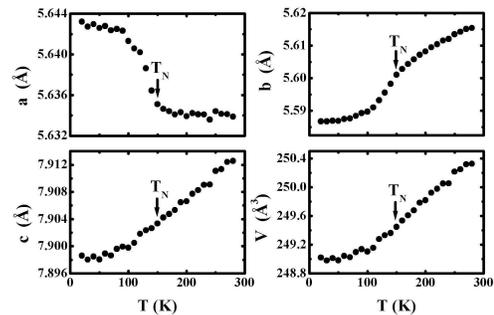}%
\caption{  Temperature dependence of the orthorhombic lattice
constants and of the lattice volume obtained by x-ray diffraction.} 
\label{fig:2}
\end{figure}

The anomalous temperature dependence of the $a$ and $b$ lattice parameters
is not restricted to  a temperature range close
to \tn , see Fig. 3. 
In particular the $a$-parameter which is almost constant between
200-500\ K deviates from a normal thermal expansion over a large
temperature interval.
Here, it is instructive to analyze the orthorhombic splitting.
The lattice in $Pbnm$ results from that of an ideal cubic perovskite 
(denoted by subscript $c$)
by setting $a =\sqrt{2} a_c$, $b =  \sqrt{2} a_c$ and $c =  2a_c$.
Considering a rigid tilt of the ideal octahedron one obtains the non-equalities
$c \ne {1 \over \sqrt{2}} (a + b  )$
due to the rotation and the tilt and $a \ne b$
due to the tilt distortion only.
However, tilting of an ideal octahedron around the $b$-axis would lead
to $b > a$, whereas \latio ~ exhibits just the opposite behavior
below about 650\ K.
The sign of the orthorhombic distortion,
$ \epsilon = { b - a \over a +b}$, already indicates
that the octahedron shape in \latio ~ is not ideal.

\begin{figure}[t]

\includegraphics*[width=0.5\columnwidth]{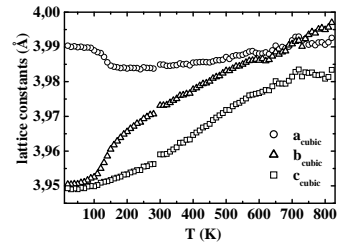}%

\caption{  Temperature dependence of the orthorhombic lattice
constants scaled to the cubic perovskite lattice (x-ray results).
} \label{fig:3}
\end{figure}

We have determined the crystal structure
on our almost stoichiometric sample using  
high resolution neutron diffraction.
The results are given in table 1.
The splitting in the
Ti-O-bond distances is small though not negligible.
There is, however, a large splitting in the O2-O2 edge lengths of 
the octahedron basal plane, which has escaped previous attention.
At 8\ K the edges along $a$ and $b$ differ by about 4\% .
This distortion is comparable in its strength
to that observed for the Ti-O-bond distances in YTiO$_3$ 
which is considered as being orbitally ordered \cite{imada}.
In \latio , the octahedron elongation
perpendicular to the tilt axis
over-compensates the shortening of $a$ through a rigid tilt.
The negative sign of the orthorhombic strain in \latio ~ is hence caused
by the elongation of the octahedron along $a$.
Interestingly, this effect is found to be strengthened at the
antiferromagnetic transition; the behavior of the O2-O2 bond distances
accounts for the anisotropic anomalies : an elongation along $a$ and
a shrinking along $b$.
Comparing our new results with those of ref. \cite{eitel,maclean} 
one recognizes that
the splitting in Ti-O-bond lengths, the splitting in the
O2-O2 edges as well as the anomalies near $T_N$ 
are less pronounced in the samples of ref. \cite{eitel,maclean}.
Non-stoichiometry seems to significantly
reduce the long-range distortion of the TiO$_6$-octahedron.
This was confirmed by our
own studies on samples with excess oxygen which revealed a continuous 
reduction of the orthorhombic strain and the anomalies around $T_N$.

\begin{figure}[t]

\includegraphics*[width=0.550\columnwidth]{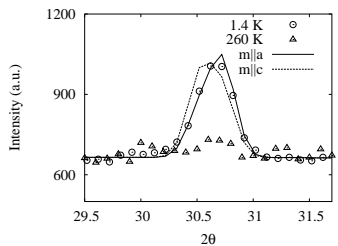}%

\caption{  Part of the neutron diffraction pattern for different
temperatures, the two lines correspond to Rietveld fits with
the antiferromagnetic component either along $a$ or $c$.}
\label{fig:4}
\end{figure}

Meijer et al. performed neutron diffraction studies on single
crystals in order to determine the magnetic structure \cite{meijer}. 
According to the symmetry of the GdFeO$_3$ structure \cite{bertaut,meijer}
a G-type antiferromagnetic order may be associated with a weak
ferromagnetic moment. However, Meijer et al. could not
distinguish between the following two configurations due to twinning : 
either a G-type moment along $a$ and ferromagnetism along $c$ or a G-type
moment along $c$ and ferromagnetism along $a$. Using the high flux
diffractometer data we get evidence that the ordered
antiferromagnetic moment points along the $a$-direction, see Fig. 4,
although some underlying weak nuclear reflection complicates the analysis. 
Constraining the crystal structure to the high resolution results given in
Table I,
we find a low temperature ordered moment of 0.57(5) \muB ~ 
slightly higher than previous studies.

\begin{figure}[t]
\includegraphics*[width=0.97\columnwidth]{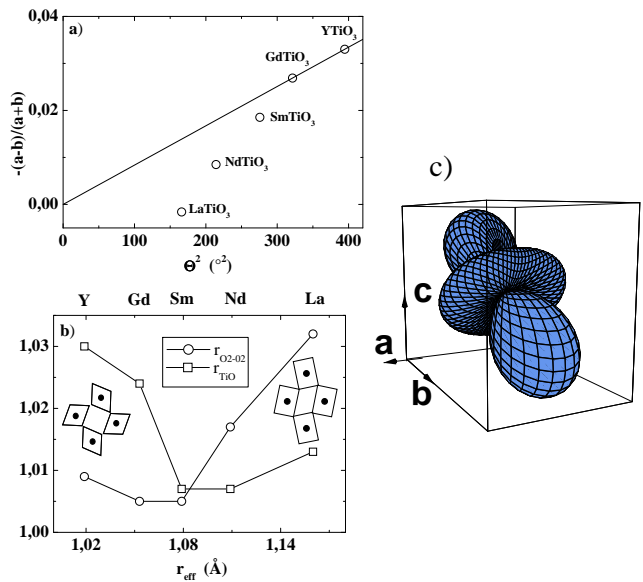}%
\caption{  a) Plot of the orthorhombic distortion against the
square of the tilt angle for different RETiO$_3$. b) Plot of the
two prominent distortions of the TiO$_6$-octahedron against the
ionic radius of the RE in RETiO$_3$. Left scale denotes the ratio
of the longest to shortest TiO-bond length, right scale the ratio
of the O2-O2 basal plane edge lengths. Except our \latio ~
data were taken from ref. \cite{maclean}. The insets in b)
show a drawing of the two different octahedron distortion schemes.
c) Shape of the \tzg -orbital in \latio , which is lowered in energy due 
to the distortion of the octahedron (axes correspond to the
$Pbnm$ lattice).}
\label{fig:5}
\end{figure}

The crystal structure within the  RETiO$_3$ series
is determined through the ionic radius of the RE. 
A  smaller radius implies increasing tilt and rotation distortions.
In Fig. 5a) we plot the orthorhombic splitting against the
square of the tilt angle. There is a clear deviation from
the proportionality expected in Landau-theory \cite{bruce}, 
which we attribute to the anomalous stretching of the octahedron basal plane.
Without this anomalous elongation one would expect an orthorhombic
splitting in \latio ~ three times larger and of positive sign.
The anomalous octahedron elongation at the magnetic
ordering is, therefore, about an order of magnitude smaller than the total
deformation in \latio , see Fig. 3 and table I.
In Fig. 5b) we compare the octahedron deformation in \latio ~ 
with that seen in the RETiO$_3$-series using the data of MacLean et 
al. \cite{maclean}.
In \latio ~ we find a small variation in the TiO-bond distances,
described by $r_{TiO}={Ti-O_{long}\over Ti-O_{short}}$, but a strong
variation in the basal plane edge lengths described by
$r_{O2-O2}={O2-O2_{long}\over O2-O2_{short}}$.
In contrast, in YTiO$_3$ the variation in the Ti-O-distances is
strong and that in the O2-O2-lengths is small, as it was recently confirmed
\cite{heester}. 
Fig. 5b) shows that these distortions change continuously in the RETiO$_3$ series.
The two distinct deformations of the octahedra imply different
orbital ordering schemes. The variation of the Ti-O-distances
arises from the shift of an O2-site  towards a Ti-site;
in  consequence, the distance to the opposite Ti is elongated
and antiferroorbital-type ordering results which has been
directly observed in YTiO$_3$  \cite{akimitsu}.
In contrast the elongation of the basal planes along $a$ in \latio ~
implies ferro-orbital-type ordering.
Following Goodenough-Kanamori rules these orderings imply
ferromagnetic and antiferromagnetic coupling, respectively.
Most interestingly, the cross-over between the two types of structural
deformation \cite{note-struc} coincides with the cross-over from 
antiferromagnetic to
ferromagnetic ordering \cite{greedan,amow1,amow2} in the RETiO$_3$-compounds.
With decreasing RE ionic radius
NdTiO$_3$ and SmTiO$_3$ exhibit still antiferromagnetic order
with small $T_N$
\cite{amow1,amow2}, whereas GdTiO$_3$ is ferromagnetically ordered.
A similar magnetic crossover is seen in the La$_{1-x}$Y$_x$TiO$_3$-series 
\cite{okimoto}.
This coincidence between the structural and magnetic crossovers 
clearly documents the  relevance of the octahedron deformation
for the magnetism in RETiO$_3$.

%

In the following we analyze the effect of the octahedron basal plane 
elongation  on the $t_{2g}$-energy levels.
Qualitatively one may argue, that the octahedron elongation splits 
the degenerate $xz$ and $yz$-levels into ${1\over \sqrt{2}}(xz+yz)$ 
and ${1\over \sqrt{2}}(xz-yz)$.
In the notation used here
the orbital coordinates refer to the cubic perovskite lattice,
i.e.  $x, y, z$ are parallel to the Ti-O-bonds.
We have calculated the $t_{2g}$-level splitting within a
full Madelung-sum point charge model taking into account the
second order covalent contribution.
Using the experimentally determined crystal structure, 
we obtain a sizeable splitting: one level
is about 0.24eV below the two others which remain almost degenerate.
The crystal field splitting is about one order of magnitude stronger
than the spin orbit coupling, and we have to conclude that the 
$t_{2g}$-orbitals are not degenerate in \latio ~; but the splitting is not
sufficiently strong to completely quench the orbital moment.
The occupied orbital can be constructed as a linear combination of the
three standard $t_{2g}$-orbitals~: 
$0.77({1\over \sqrt 2}\vert xz+yz >) \pm 0.636 \vert xy>$. 
The admixture of the $xy$-orbital to ${1\over \sqrt 2}\vert xz+yz >$
alternates in its sign along the $c$-axis and is constant in the $a,b$-plane.
The occupied orbital is oriented in the orthorhombic $b,c$-plane and
has almost the shape of the usual $3z^2-r^2$
orbital pointing near the cubic [111]$_c$ direction \cite{comment}
(56$^o$ angle  with the $c$-axis compared to 54.7$^o$ for the 
[111]$_c$-direction), see Fig. 5c). 
The lowered orbital points to the center of one of the octahedron
triangles-faces 
which are formed by two O2's and one O1. 
It is interesting to note that the O2-O1 distances
of this triangle are elongated compared to the average. 
This means that the TiO$_6$-octahedron is 
compressed along the direction of the occupied orbital.
Since the occupied orbital is not oriented along one of the bonds the resulting 
antiferromagnetic exchange is not expected to be very anisotropic.
Assuming the calculated orbital splitting and quantum fluctuations we obtain 
an ordered moment of 0.72 \muB .  The ordered moment will be further reduced
as covalence  will diminish the level splitting, admix orbital
contributions and  transfer a part of the spin to the oxygen-sites 
which are not active in the G-type antiferromagnetic ordering. 
Similar to the proposal by Moshizuki and Imada \cite{moshizuki}
we may hence explain most of the magnetic properties of \latio ~
basing on the structural distortion observed in the almost stoichiometric
compound.
The distortion assumed in reference \cite{moshizuki} has
not the same geometry as the one we find but is comparable in magnitude.

Since \latio ~ exhibits intermediately strong \tzg -level splitting, 
the orbital occupancy,
the admixture of an orbital moment
and the resulting antiferromagnetic exchange interaction 
depend sensitively on the crystal structure. 
On one side the system may gain magnetic energy in the ordered phase
either by an enhancement of the magnetic interaction  parameter 
or via spin-orbit coupling. 
On the other side there will be a loss of elastic energy
associated with the additional structural distortion.
In the ordered state there should be a trend to 
increase the orbital ordering in agreement with experiment;
according to our calculation the level splitting is enhanced by
6\% between 155 and 8\ K.

In summary we have found a sizeable deformation 
of the TiO$_6$-octahedron in \latio ~ which causes a relevant
splitting in the \tzg -levels (about 0.24eV).

We gratefully acknowledge discussions with M. Gr\"uninger, G. Khaliullin 
and M. Laad.
This work was supported by the Deutsche Forschungsgemeinschaft through SFB 608.


\begin{thebibliography}{}
\bibitem{note-struc}Caution has to be taken concerning the stoichiometry 
of the compounds studied in \cite{maclean}; 
also, the early magnetic study \cite{greedan} 
did not observe the antiferromagnetism in  NdTiO$_3$ and SmTiO$_3$, structural
data on these samples indicate a smoother crossover.
\bibitem{comment} In a rotated coordinate system the orbital 
can be described by $3z^2-(1.1x^2+0.9y^2+0.8z^2)$.
\end{thebibliography}
\end{document}